\documentclass[journal]{IEEEtran}
\RequirePackage{etex}
\usepackage{myStyleIEEE}
\usepackage{ushort}
\IEEEoverridecommandlockouts
\def\Z{\mathbb{Z}}
\def\R{\mathbb{R}}
\def\C{\mathbb{C}}
\usepackage{siunitx}
\AtBeginDocument{\DeclareSIUnit{\MWh}{MWh}}
\usepackage{mathtools}
\usepackage{isomath}
\let\underbrace\LaTeXunderbrace

\begin{document}

\title{MPC-Based Fast Frequency Control of Voltage Source Converters in Low-Inertia Power Systems}

\renewcommand{\theenumi}{\alph{enumi}}

\newcommand{\uros}[1]{\textcolor{magenta}{$\xrightarrow[]{\text{U}}$ #1}}
\newcommand{\vaggelis}[1]{\textcolor{blue}{$\xrightarrow[]{\text{V}}$ #1}}
\newcommand{\petros}[1]{\textcolor{red}{$\xrightarrow[]{\text{P}}$ #1}}
\newcommand{\ognjen}[1]{\textcolor{pPurple}{$\xrightarrow[]{\text{O}}$ #1}}

\author{Ognjen~Stanojev,~\IEEEmembership{Student~Member,~IEEE,}
        Uros~Markovic,~\IEEEmembership{Member,~IEEE,}
        Petros~Aristidou,~\IEEEmembership{Senior~Member,~IEEE,}
        Gabriela~Hug,~\IEEEmembership{Senior~Member,~IEEE,} 
        Duncan~Callaway,~\IEEEmembership{Senior~Member,~IEEE,} 
        Evangelos~Vrettos,~\IEEEmembership{Member,~IEEE}\vspace{-0.35cm}
}

\maketitle
\IEEEpeerreviewmaketitle

\begin{abstract}
A rapid deployment of renewable generation has led to significant reduction in the rotational system inertia and damping, thus making frequency control in power systems more challenging. This paper proposes a novel control scheme based on Model Predictive Control (MPC) for converter-interfaced generators operating in a \textit{grid-forming} mode, with the goal of exploiting their fast response capabilities to provide fast frequency control service to the system. The controller manipulates converter power injections to limit the frequency nadir and rate-of-change-of-frequency after a disturbance. Both \textit{centralized} and \textit{decentralized} MPC approaches are considered and compared in terms of performance and practical implementation. Special attention is given to the decentralized controller by generating an explicit MPC solution to enhance computational efficiency and reduce hardware requirements. Simulation results obtained from a high-fidelity differential-algebraic equation model of the IEEE 39-bus system demonstrate the effectiveness of the proposed control schemes.
\end{abstract}


\vspace{-0.2cm}
\section{Introduction} \label{sec:intro}

\lettrine[lines=2]{A}{large-scale} integration of converter-interfaced generation imposes new challenges on real-time power system control and operation, as the lack of rotational inertia and governor droop control (i.e., damping) leads to faster dynamics and larger frequency deviations \cite{Milano2018}. In order to mitigate potential stability issues and improve the resilience of low-inertia systems, new ancillary services such as Fast Frequency Control (FFC) are needed~\cite{EirGrid2012}. These requirements can be fulfilled by grid-forming (i.e., grid-supporting) Voltage Source Converters (VSCs) and the associated DC-side energy buffers, as they can effectively adjust the power output in response to frequency deviations. 

The two most common \textit{grid-forming} VSC control approaches in the literature are a Virtual Synchronous Machine (VSM), i.e., an emulation technique based on the swing dynamics of a synchronous machine \cite{Zhong2011}, and a droop-based control which takes advantage of the traditional droop characteristic for regulating the converter's active and reactive power output \cite{UrosGM}. Nevertheless, the majority of proposed control strategies focuses solely on the converter's AC-side, disregarding the DC-link dynamics in the process and making the simplifying assumption that an infinite amount of power and energy is available at the DC-side capacitor \cite{Ashabani2014}. Moreover, while specifying a constant droop gain leads to satisfactory VSC performance under small frequency deviations, it prevents the converter from utilizing its maximum power capacity in emergency cases. 

Model Predictive Control (MPC), an optimization-based, discrete-time control scheme, appears to be promising for incorporating all of the aforementioned aspects into a uniform problem formulation \cite{MPCbook}. The capability to compute optimal control inputs based on predictions of future state evolution using a state-space system model and disturbance forecasts, while taking operational constraints into consideration, has made MPC attractive for frequency regulation in power systems.

In recent years several studies have considered the application of MPC in Automatic Generation Control (AGC). Centralized \cite{Ersdal2016}, hierarchical \cite{Shiroei2014} and distributed \cite{Venkat2008} approaches have been proposed and shown to improve frequency regulation and robustness to uncertainty when compared to standard PI control. In contrast, only a few studies have addressed the application of MPC to fast frequency control \cite{Ulbig2013,Papangelis2018,Fuchs2014}. A real-time optimal control scheme based on \textit{explicit} MPC for regulating frequency and providing inertial response was presented in \cite{Ulbig2013}. Although the advantages of an explicit MPC scheme in fast frequency regulation were illustrated, this study used a simplified power system model and did not include converter-based generation.

The drawbacks of the aforementioned study were addressed in \cite{Papangelis2018} and \cite{Fuchs2014}, where MPC-based frequency support through HVDC grids was investigated. In \cite{Papangelis2018}, a \textit{decentralized} MPC control scheme for frequency containment in emergency situations was proposed. Frequency predictions are made based on Rate-of-Change-of-Frequency (RoCoF) measurements and the VSC output is adjusted if constraint violations are detected or expected. Despite being decentralized, this approach requires global information about the grid topology and HVDC converter locations to calculate sensitivity factors corresponding to DC-voltage droop. Tuning of such parameters as well as the increased computational burden (due to solving the MPC problem online) are the limitations of this approach. Alternatively, stabilization of large power systems using VSC-based HVDC links equipped with a \textit{centralized} MPC controller was analyzed in \cite{Fuchs2014}. Based on global measurements, the VSC injections are manipulated to damp out oscillations in the system. However, fast and reliable communication links are required to leverage MPC benefits and resolve potential stability issues arising from communication delays and failures.

This paper presents both a centralized and a decentralized MPC-based FFC strategy that can be incorporated as an additional layer to the primary frequency control (droop or VSM-based). While not active in normal operation, the MPC is triggered in case of large disturbances to keep the frequency deviation and RoCoF within limits prescribed by the operator. We start by introducing improvements to the frequency prediction in \cite{Papangelis2018} by employing a Center-of-Inertia (CoI) frequency dynamics model of a low-inertia system developed in \cite{UrosLQR}. Subsequently, model identification methods are applied to estimate the parameters of the frequency response model based on historical data. Furthermore, improvements in computational efficiency of the decentralized MPC approach by means of an offline explicit solution scheme are assessed. Finally, in contrast to the studies in \cite{Ulbig2013,Papangelis2018,Fuchs2014}, the proposed control design is verified through time-domain simulations using a detailed Differential Algebraic Equation (DAE) model of a low-inertia system described in \cite{UrosDAE}.

The rest of the paper is structured as follows. In Section~\ref{sec:MPC_FFC}, a general overview of MPC application to FFC is provided and the MPC-based supervisory layer is introduced in the converter control scheme. Sections~\ref{sec:decentralized} and \ref{sec:centralized} elaborate on the design of decentralized and centralized controllers, respectively, as well as the underlying prediction models. Additionally, in Section~\ref{sec:decentralized}, an explicit MPC solution and the model identification procedure for estimating the prediction model parameters are presented. Simulation results from different case studies are illustrated in Section~\ref{sec:res}, whereas Section~\ref{sec:conclusion} draws the main conclusions and discusses future work.

\section{MPC-Based Fast Frequency Control for VSCs} \label{sec:MPC_FFC}

\subsection{MPC Application to Fast Frequency Control}

Traditionally, primary frequency control together with system's rotational inertia was sufficient for containing frequency excursions in emergency cases. However, as the system inertia and hence the time constants of frequency dynamics decrease, the primary control response times fail to meet the requirements for maintaining the frequency within limits in the immediate aftermath of a disturbance. This raises the need for control schemes operating on shorter timescales \cite{Terzija2018}; a service that could be ideally provided through rapid active power delivery of the VSC interfacing the renewable generation or battery storage unit to the network. 

The basic MPC concept can be outlined as follows. At the current discrete time step $k\in\Z_{\geq0}$, the controller receives the latest available measurements and uses state-space-based predictions to compute the optimal control sequence $u^*(k),u^*(k+1),\dots,u^*(k+N-1)$ over a horizon of $N\in\Z_{\geq0}$ future time steps to satisfy the required constraints at the minimum cost. Subsequently, only the control action for the first time sample is applied to the system and the rest of the sequence is discarded. The procedure is repeated for every following sample time step with the inclusion of updated process measurements.

An MPC-based FFC scheme for converter-interfaced generators can be developed according to the following approach. After a large disturbance, sufficient information is promptly collected by observing changes in system variables in order to predict frequency evolution for the next time period. Optimal control actions are then computed based on the state-space predictions to prevent critical threshold violations (e.g., frequency nadir or RoCoF), while respecting device-level constraints. Subsequently, each converter unit participating in FFC attempts to counteract a part of the estimated disturbance. The frequency control scheme needs to be compatible with and complementary to all grid-forming controllers and is therefore designed as a supervisory control layer. Without loss of generality, this chapter focuses solely on droop-based grid-forming converters, but the same application can be easily applied to VSM operation mode based on a well-known small-signal equivalence between the two models \cite{UrosEnergyCon}. 

\subsection{VSC-Level Implementation} \label{subsec:vsc_ctrl}
\begin{figure}[!t] 
	\centering
	\scalebox{0.834}{\includegraphics[]{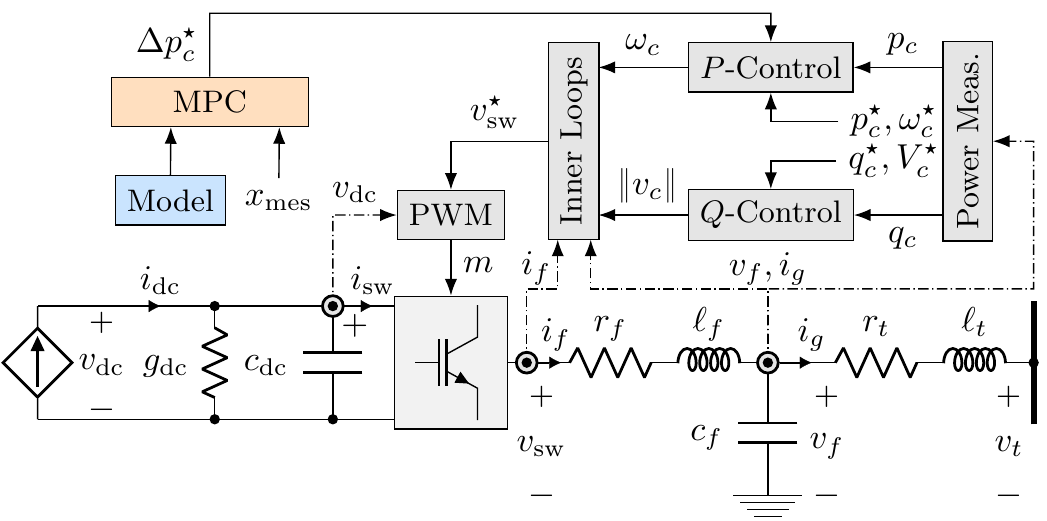}}
	\caption{Simplified diagram of the implemented control structure with the MPC-based supervisory layer. }
	\label{fig:vsc_ctrl_new}
\end{figure}

The model of a VSC used in this work comprises a DC-side circuit, an AC-side circuit and a lossless switching unit which modulates the DC-capacitor voltage $v_\mathrm{dc} \in \R_{>0}$ into an AC voltage $v_{\mathrm{sw}} \in \R^2$, as depicted in Fig.~\ref{fig:vsc_ctrl_new}. 
Modeling and control of the converter is implemented in a Synchronously-rotating Reference Frame (SRF), with the mathematical model defined in $dq$-vector form $x\coloneqq (x^d,x^q)\in\R^2$ and per-unit. Hence, the electrical subsystem including an RLC filter $(r_f, \ell_f, c_f) \in \R^3$ and a transformer equivalent $(r_t, \ell_t) \in \R^2$ can be represented by
\begin{subequations}\label{eq:VSC_spaceVect}
\begin{align}
\dot{i}_f&=\frac{\omega_{b}}{\ell_f}(v_\mathrm{sw}-v_f)-\left(\frac{r_f}{\ell_f}\omega_{b}+j\omega_{b} \omega_r\right)i_f, \label{eq:elSys1} \\
\dot{v}_f&=\frac{\omega_{b}}{c_f}(i_f-i_g)-j\omega_b \omega_r v_f, \\
\dot{i}_g&=\frac{\omega_b}{\ell_t}(v_f-v_t)-\left(\frac{r_t}{\ell_t}\omega_b+j\omega_b \omega_r\right)i_g, \label{eq:elSys3}
\end{align}
\end{subequations}
where $i_f \in \R^2$ and $v_f \in \R^2$ are the filter current and voltage, $i_g \in \R^2$ denotes the transformer current, and $v_t\in\R^2$ is the voltage at the connection terminal; the system base frequency is represented by $\omega_b\in\R_{>0}$ and $\omega_r\in\R_{>0}$ is the normalized reference for the angular velocity of the $dq$-frame.
 
The outer control loop consisting of active and reactive power controllers provides output voltage magnitude $\|v_c\|\in\R$ and frequency $\omega_c\in\R$ references by adjusting the predefined setpoints $(p_c^\star,\omega_{c}^\star,q_c^\star, V_c^\star) \in \R^4$ according to the droop control law and the power measurements $p_c\coloneqq v^\mathsf{T}_f i_g$ and $q_c \coloneqq v^\mathsf{T}_f j^\mathsf{T} i_g$: 
\begin{subequations}
\begin{align}
    &\omega_{c} \coloneqq \omega_{c}^{\star} + R_{c}^p (p_{c}^{\star}+\Delta p_{c}^{\star}-\tilde{p}_c),
   &&\dot{\tilde{p}}_c \coloneqq \omega_f (p_c-\tilde{p}_c), \label{eq:w_c} \\
    &\|v_{c}\| \coloneqq V_{c}^{\star} + R_{c}^q (q_{c}^{\star}-\tilde{q}_c), 
   &&\dot{\tilde{q}}_c \coloneqq \omega_f (q_c-\tilde{q}_c),
\end{align}
\end{subequations}
with $R_{c}^p \in \R_{\geq0}$ and $R_{c}^q \in \R_{\geq0}$ denoting the active and reactive power droop gains, $\tilde{p}_c \in \R$ and $\tilde{q}_c \in \R$ representing the low-pass filtered active and reactive power measurements, $\omega_f \in \R_{\geq0}$ being the low-pass filter cut-off frequency and $\Delta p_c^{\star} \in \R$ indicating the setpoint change generated by the supervisory layer. Assuming constant $\Delta p_c^{\star}$, the RoCoF state $\dot{\omega}_c \in \R$ can be computed from \eqref{eq:w_c} as 
\begin{equation}
    \dot{\omega}_c = R_{c}^p \omega_f (\tilde{p}_c-p_c). \label{eq:rocof_state}    
\end{equation}
The output of active and reactive power controllers is then passed to the cascade of voltage and current controllers (so-called inner control loop), computing a switching voltage reference $v_\mathrm{sw}^\star\in\R^2$. It encompasses a PI voltage controller
\begin{subequations} \label{eq:srf_v}
\begin{align} 
    \dot{\xi} &=  v^\star_f - v_f, \\
    i_f^\star &= K_P^v (v^\star_f - v_f) + K_I^v \xi + K_F^v i_g + j\omega_c c_f v_f, \label{eq:srf_v_b}
    \end{align}
\end{subequations}
that provides a reference $i_f^\star\in\R$ for a current PI controller 
\begin{subequations}\label{eq:srf_i}
\begin{align} 
    \dot{\gamma} &= i_f^\star - i_f,\\
    v_\mathrm{sw}^\star &= K_P^i (i_f^\star - i_f) + K_I^i \gamma + K_F^i v_f + j\omega_c \ell_f i_f, \label{eq:srf_i_b}
    \end{align}
\end{subequations}
where $(K_P^v,K_P^i)\in\R^2_{>0}$, $(K_I^v,K_I^i)\in\R^2_{\geq0}$ and $(K_F^v,K_F^i)\in\mathbb{Z}_{\{0,1\}}^2$ are the respective proportional, integral, and feed-forward gains, $\xi\in\R^2$ and $\gamma\in\R^2$ represent the integrator states, and superscripts $v$ and $i$ denote the voltage and current controllers respectively. Finally, we assume that the modulation voltage reference $v_\mathrm{sw}^\star$ is perfectly transformed to the AC side, i.e., $v_\mathrm{sw}\coloneqq v_\mathrm{sw}^\star$.

The DC-side model includes a battery storage unit with the energy capacity $e_b \in \R$, interfaced with the converter through a parallel connection of the capacitance $c_\mathrm{dc}\in\R_{>0}$ and the conductance $g_\mathrm{dc}\in\R_{>0}$. The underlying dynamics are described by
\begin{subequations}\label{eq:model_DCside}
\begin{align}
c_\mathrm{dc}\omega_b^{-1} \dot{v}_\mathrm{dc} &= -g_\mathrm{dc} v_\mathrm{dc} -i_\mathrm{sw}+i_\mathrm{dc}, \label{eq:model_vdc}\\
\dot{\chi} &\coloneqq (p_{\mathrm{dc}} - p_\mathrm{sw})e_b^{-1},
\end{align}
\end{subequations}
where $i_\mathrm{sw}$ denotes the current flowing into the switching block and $i_\mathrm{dc}$ is the net current of the battery and the renewable generation represented by the DC-current source. The battery State-of-Charge (SoC) $\chi\in\R_{\geq0}$ is derived based on the power balance between the converter's DC-side input power $p_{\mathrm{dc}} \coloneqq v_{\mathrm{dc}}i_{\mathrm{dc}}\in\R$ and the AC-side output power $p_\mathrm{sw}\coloneqq v_\mathrm{sw}^\mathsf{T}i_f$ before the filter.

Finally, a PI controller is employed to track the DC-voltage setpoint $v^\star_\mathrm{dc} \in \R_{>0}$ by adjusting the DC-current source
\begin{subequations}\label{eq:DCPI2}
\begin{align}
    \dot{\chi} &= v^\star_\mathrm{dc} - v_\mathrm{dc}, \label{eq:DCPI}\\
    i_\mathrm{dc} &=K_P^\mathrm{dc} (v^\star_\mathrm{dc}-v_\mathrm{dc}) + K_I^\mathrm{dc} \chi + K_F^\mathrm{dc} i^\star_\mathrm{dc}, \label{eq:DCPI1}
\end{align}
with $\chi\in\R$ being the internal state variable, and proportional, integral, and feed-forward gains denoted by $K_P^\mathrm{dc} \in \R_{>0}$, $K_I^\mathrm{dc} \in \R_{\geq 0}$, and $K_F^\mathrm{dc} \in \{0,1\}$, respectively. The DC current reference $i^\star_\mathrm{dc}\in\R_{>0}$ at a nominal operating point $(V_c^\star,p_c^\star,q_c^\star)$, including DC and AC circuit losses, is given by 
\begin{equation}
i^\star_\mathrm{dc} \coloneqq {v^\star_\mathrm{dc}}^{-1} \left(p_c^\star+r_f \frac{{p_c^\star}^2+{q_c^\star}^2}{{V_c^\star}^2}\right) + g_\mathrm{dc} v^\star_\mathrm{dc}, \label{eq:dc:ctrl}
\end{equation}
\end{subequations}
which indicates that for $v_\mathrm{dc}=v_\mathrm{dc}^\star$ the DC-side current will be $i_\mathrm{dc}=i_\mathrm{dc}^\star$.

The supervisory control layer employs an MPC which, based on a frequency prediction model and the newest available measurements $x_\mathrm{mes}\in\R^{m}$, generates a signal $\Delta p_c^{\star}$ to modulate the active power setpoint in response to a disturbance. Whereas inactive during normal operation, the supervisory layer is triggered in emergency cases and remains active until the new steady state is reached.

\section{Decentralized Control Design} \label{sec:decentralized}

The main goal of the decentralized approach is to design an FFC scheme where each VSC relies solely on local measurements and proportionally participates in disturbance mitigation based on its location in the system. Optimally, the converters closer to the fault shall provide more support in order to avoid stress on the transmission lines and losses. Communication among converters is avoided and each individual VSC can be included in FFC support in a \textit{plug-and-play} fashion.

Since there is no need for provision of FFC in normal operation, the controller remains inactive until a disturbance is detected. As a disturbance indicator, internally obtainable RoCoF estimates defined by \eqref{eq:rocof_state} are used. As long as the RoCoF stays within a predefined deadband the controller remains idle. Once the threshold is violated, the FFC is activated and kept in operation until the average RoCoF returns below prescribed margins. A benefit of such approach lies in the simultaneous activation and synchronous action of all VSCs participating in FFC, without the need for communication and independent of converter location.

In the remainder of this section, the derivation of an accurate frequency prediction model is presented together with mathematical formulation of the decentralized MPC problem. Moreover, a data-driven approach for estimating the prediction model parameters is described as well as the explicit MPC solution scheme for efficient computation of the optimal control inputs.

\subsection{Frequency Prediction Model}
The work in \cite{Papangelis2018} proposes a simple RoCoF-based frequency prediction model, where at each discrete time step $k\in\Z_{\geq0}$ an instantaneous RoCoF measurement $r_f(k)\in\R$ is obtained and used to estimate the frequency deviation $\Delta f(k+j)\in\R$ for $N \in\Z_{\geq0}$ future time steps $j\in\{1,2,\dots,N\}$ of the prediction horizon, as follows:
\begin{equation} \label{eq:papangelis_df}
    \Delta f(k+j) = r_f(k) T_s + \frac{\Delta p_c(k+j)}{2H}T_s.
\end{equation}
Here, $T_s\in\R_{>0}$ denotes the length of a single time step, $H\in\R_{>0}$ is the aggregate inertia constant, and $\Delta p_c(k+j)\in\R$ represents the VSC power adjustment at the respective time step. Although simple and convenient for MPC implementation, the proposed model predicts a linear frequency decay based on the instantaneous RoCoF at the onset of the disturbance, and hence leads to large errors when estimating the frequency nadir.  

We improve the prediction accuracy by employing a CoI frequency model of a generic low-inertia system introduced in \cite{UrosLQR}, accounting for the inertial response and primary frequency control of SGs as well as the frequency support of converter-based generators. In Laplace domain, it can be represented by a simplified, yet sufficiently accurate, transfer function $G(s)$ relating the CoI frequency deviation $\Delta f(s)\in\C$ to a change in power $\Delta p(s)\in\C$
\begin{equation}
    G(s) = \frac{\Delta f(s)}{\Delta p(s)} = \frac{1}{MT}\frac{1+sT}{s^2+2\zeta\omega_n s + \omega_n^2}. \label{eq:G}
\end{equation}
The natural frequency $\omega_n\in\R_{>0}$ and damping ratio $\zeta\in\R_{>0}$ are computed as
\begin{equation}
    \omega_n = \sqrt{\frac{D+R_g}{MT}}, \quad \zeta = \frac{M+T(D+F_g)}{2\sqrt{MT(D+R_g)}}, \label{eq:wn}
\end{equation}
with parameters $M\in\R_{>0}$ and $D\in\R_{>0}$ representing the weighted system averages of inertia and damping constants, respectively. Similarly, $R_g\in\R_{>0}$ and $F_g\in\R_{>0}$ denote the average inverse droop control gain and the fraction of total power generated by the high-pressure turbines of Synchronous Generators (SGs), while $T\in\R_{>0}$ stands for the generator time constant. A simplification of assuming equal time constants for all SGs is made according to the analysis in \cite{Ahmadi2014}, suggesting that the frequency nadir and RoCoF are the least sensitive metrics to turbine time constants. Moreover, the inverter time constants are
approximately 2-3 orders of magnitude lower than the ones
of synchronous machines. A verification of the proposed frequency model can be found in \cite{UrosLQR}, together with definitions and analytic expressions of all relevant system parameters.

Transfer function \eqref{eq:G} can now be transformed into a controllable canonical state-space model
\begin{subequations}
\label{eq:ss_general}
\begin{align} 
\underbrace{
\begin{bmatrix}
    \dot{q}_1(t) \\ 
    \dot{q}_2(t)
\end{bmatrix}
}_{\dot{x}(t)}
&= 
\underbrace{
\begin{bmatrix}
    0 & I \\
    -\omega_n^2 & -2 \zeta \omega_n
\end{bmatrix} 
}_{A}
\begin{bmatrix}
    q_1(t) \\ 
    q_2(t)
\end{bmatrix}
+
\underbrace{
\begin{bmatrix}
    0 \\ I
\end{bmatrix}
}_{B}
\Delta p(t), \\
\Delta f(t) &= 
\underbrace{
\begin{bmatrix}
    \frac{1}{MT} & \frac{1}{M}
\end{bmatrix}
}_{C}
\begin{bmatrix}
    q_1(t) \\ 
    q_2(t)
\end{bmatrix},
\end{align}
\end{subequations}
where $A\in\R^{2\times2}$, $B\in\Z_{\geq0}^{2}$ and $C^\mathsf{T}\in{\R_{\geq0}^2}$ denote the state-space matrices, and $x \coloneqq (q_1,q_2)\in\R^2$ represents the state vector that does not correspond to any physical variables in the system. A zero-order hold equivalent of the state-space model is used to obtain a discrete-time form suitable for MPC application.

Since $\Delta p(t)$ is a control input in \eqref{eq:ss_general}, it is necessary to measure the disturbance signal prior to predicting the frequency evolution. By applying a stepwise disturbance $\Delta p(s)= \Delta P/s$ to the model in \eqref{eq:G}, a relationship between the maximum instantaneous RoCoF (i.e., RoCoF in the immediate aftermath of the disturbance, before any system controls are activated) $\dot{\omega}_\mathrm{max}\in\R$ and the disturbance magnitude $\Delta P\in\R$ is known, and yields $\dot{\omega}_\mathrm{max} \coloneqq -\Delta P/M$ \cite{UrosLQR}. Note that the formulation is presented in per-unit, i.e., $\dot{\omega}_\mathrm{max}=\dot{f}_\mathrm{max}/f_b$. Considering that the RoCoF measurements are internally available at each grid-forming VSC, the magnitude of the system disturbance can be locally estimated and subsequently used for frequency evolution prediction in \eqref{eq:ss_general}.

\subsection{Decentralized MPC Formulation}
Let us denote by $\mathcal{H}=\{k, k+1, \dots,k+N\}$ the MPC prediction horizon of length $N$, including $k$ as the current time step. The proposed optimization problem aims at minimizing the total control effort over the full horizon, i.e., $\forall k\in\mathcal{H}$, as follows:
\begin{subequations}
\label{eq:optimization1}
\begin{alignat}{3}
    &\underset{u}{\min} \quad && \sum_{k\in\mathcal{H}} C_P(k)  \lVert\Delta p_c^\star(k) \rVert + C_H \left( \lVert\eta_{f} \rVert_\infty +  \lVert\eta_{r} \rVert_\infty \right)\label{eq:objective1} \\
    &\mathrm{s.t.} \quad && x(k+1) = A_d x(k) + B_d(\Delta p_c^\star(k) + \Delta P), \label{eq:states_pred} \\
    &  && f(k) = C_d x(k) + f_0, \label{eq:otputs_pred} \\
    &  && \dot{f}(k) = \frac{f(k)-f(k-1)}{T_s}, \label{eq:rocof_discrete} \\
    & && p_c(k) = p_c^\star + \sum_{r=1}^k\Delta p_c^\star(r) + R_c^p (\omega_c^\star - \omega_c(k)),  \label{eq:power_estimated}\\
    &  && \chi(k+1) = \chi(k) + T_s\frac{p_c^\star-p_c(k)}{E_b}, \label{eq:soc_estimated}\\
    &  && \ushort{p}_{c,\mathrm{lim}} \leq p_c(k) \leq \widebar{p}_{c,\mathrm{lim}}, \label{eq:powerLimit}\\
    &  && \ushort{\chi}_\mathrm{lim} \leq \chi(k) \leq \widebar{\chi}_{\mathrm{lim}}, \label{eq:SoCLimit}\\
    &  && \ushort{f}_\mathrm{lim} \leq f(k) + R_c^p \Delta p_c^\star(k) \leq \widebar{f}_{\mathrm{lim}}, \label{eq:fciLimit}\\
    &  && \ushort{f}_\mathrm{lim} - \eta_f(k) \leq f(k) \leq \widebar{f}_{\mathrm{lim}} + \eta_f(k), \label{eq:frequencyLimit}\\
    &  && \ushort{\Dot{f}}_\mathrm{lim} - \eta_r(k) \leq \Dot{f}(k) \leq \widebar{\Dot{f}}_{\mathrm{lim}} + \eta_r(k), \label{eq:RoCoFLimit}\\
    &  && \eta_f(k) \geq 0, \eta_r(k) \geq 0, \label{eq:slackConstraints}
\end{alignat}
\end{subequations}
with $x(k)\in\R^2$ denoting a state vector at a discrete time step $k$ and $u\in\R^{N+1}$ being the vector of setpoint changes $\Delta p_c^\star(k)$. The coefficients $C_P(k)\in\R_{\geq0}$ in the objective function \eqref{eq:objective1} represent the cost of the converter action at each time step $k$. Values of the coefficients are chosen such that $C_P(k) \leq C_P(k+1)$ holds, which incentivizes the use of control resources at earlier time steps in order to prevent late reactions and frequency oscillations near the frequency limit resulting from the converter setpoint alteration. Slack variables $\eta_f\in\R_{\geq0}^{N+1}$ and $\eta_r\in\R_{\geq0}^{N+1}$, in conjunction with a large penalty factor $C_H\in\R_{>0}$, are used to relax the respective frequency and RoCoF constraints and avoid potential feasibility issues. 

The prediction model described in \eqref{eq:states_pred}-\eqref{eq:rocof_discrete} aims at anticipating the system frequency evolution for future time steps. For that purpose, the discrete-time counterpart of the frequency prediction model \eqref{eq:ss_general} is used, with $A_d\in\R^{2\times2}$, $B_d\in\Z_{\geq0}^2$ and $C_d\in{\R_{\geq0}^2}^\mathsf{T}$ describing the respective state space, $\Delta P\in\R$ denoting the estimated disturbance magnitude, $T_s\in\R_{>0}$ designating the length of a single discrete time step, and $f_0\in\R_{>0}$ representing the frequency linearization point (i.e., the nominal frequency). Equality \eqref{eq:rocof_discrete} augments the frequency model with the prediction of average RoCoF over a single time step. 

Constraints \eqref{eq:power_estimated}-\eqref{eq:fciLimit} take into account the physical limitations of the converter such as the upper and lower bounds on power output $p_c(k)$ and battery SoC $\chi(k)$. The second term in \eqref{eq:power_estimated} accumulates the setpoint changes from previous time steps and the third term accounts for the contribution of droop control; \eqref{eq:soc_estimated} is a discrete formulation of the dynamics pertaining to battery SoC, with $p_{\mathrm{dc}} = p_c^\star$. Expression \eqref{eq:fciLimit} captures the impact of droop control on system frequency, thus anticipating excessive frequency spikes coming from fast setpoint changes at the converter nodes and preventing potential converter tripping. Finally, constraints \eqref{eq:frequencyLimit}-\eqref{eq:RoCoFLimit} impose upper and lower bounds on system variables, with subscript ``$\mathrm{lim}$'' indicating the respective threshold, whereas \eqref{eq:slackConstraints} stands for trivial non-negativity constraints of slack variables.

Each VSC participating in FFC is expected to compensate for a portion of the total disturbance. Hence, the computed optimal setpoint change $\Delta p_{c_i}^\star$ of each converter $i\in\mathcal{N}_c$ is weighted by the participation coefficient $k_{p_i} \coloneqq \widebar{P}_i/P_t$ before being applied to the VSC, with $\widebar{P}_i\in\R_{>0}$ being its rated power and $P_t\in\R_{>0}$ representing the net installed power of all converters participating in FFC. 

\subsection{Model Identification}

Reliable performance of predictive control largely depends on the accuracy of the prediction model. The parameters in \eqref{eq:G}-\eqref{eq:ss_general} vary with generator dispatch changes and require information regarding the specifications of every online generator. Hence, a methodology to obtain accurate model parameters needs to be developed. Combining the known mathematical structure of the prediction model with available measurement data, the \textit{grey-box} modelling approaches can be exploited for online estimation of model parameters. 

A grey-box model is mathematically formulated as a set of continuous stochastic differential equations. It can be derived by extending the state-space model \eqref{eq:ss_general} to account for measurement errors and process uncertainty, which yields
\begin{subequations}
\label{eq:sys_ident}
\begin{align} 
    \dot{x}(t) &= A(\Omega) x(t) + B(\Omega) \Delta p(t) + \mu, \\
    \Delta f(t) &= C(\Omega) x(t) + \varepsilon,
\end{align}
\end{subequations}
with $\Omega\in\R^p$ representing the vector of unknown parameters, $\mu\in\R^2$ denoting a Wiener process and $\varepsilon\in\R$ being the measurement error. The prediction error method \cite{LjungBook} is an efficient grey-box identification approach for parameter estimation using a linear state estimator and minimizing the square of prediction residuals over all measurement samples $m \in \mathcal{M}\subset\Z_{\geq0}$. The optimization problem can be formulated as
\begin{subequations}
\label{eq:modID_optimization}
\begin{alignat}{3}
    &\underset{\Omega}{\min} \quad && \sum_{m\in\mathcal{M}}   \lVert\Delta f(m) - \Delta \hat{f}(m) \rVert_2^2  \qquad\qquad\;\, \label{eq:modID_objective}\\
    &\,\,\mathrm{s.t.} \quad  && \hat{x}(m+1) = A(\Omega)\hat{x}(m) + B(\Omega) \Delta p(m) \nonumber\\
    &  && \quad\quad \qquad+ K(\Omega) \left(\Delta f(m) - \Delta \hat{f}(m)\right), \label{eq:modID_states} \\
    &  && \Delta \hat{f}(m) = C(\Omega) \hat{x}(m), \label{eq:modID_output} \\
    &  && \hat{x}(0) = x_0, \label{eq:modID_x0} 
\end{alignat}
\end{subequations}
where $K(\Omega)\in\R^2$ is the parametrized Kalman gain, $x_0\in\R^2$ represents the initial state vector, and symbol $\hat{x}(m)\in\R^2$ denotes the vector of estimated state variables from a measurement sample $m$.

The data required for system identification process ($\Delta f(t)$ and $\Delta p(t)$ in particular) can be obtained by means of load step-change tests carried out at the converter terminal. However, to ensure observability, the disturbance magnitude needs to be significant. Another approach is to use disturbance data acquired by the operator, but would require occasional communication and result in the loss of a plug-and-play feature. Nevertheless, note that this communication will be on a much longer timescale, which preserves the controller's decentralized aspect.  The optimization problem \eqref{eq:modID_optimization} is solved using the \textsc{Matlab} System Identification Toolbox \cite{SIDtoolbox}, which also ensures stability by preserving the eigenvalues of $A-KC$ inside the unit circle.

\subsection{Explicit MPC}
Explicit MPC offers an alternative approach for computing optimal control actions without the need for executing an optimization algorithm in real time. The basis for such application lies in multi-parametric programming, whose solution yields a complete map of all optimal solutions for different operating conditions, and hence the effort needed to obtain the optimal control inputs reduces to function evaluation. The embedded control system can in turn be designed with low hardware and software requirements.

Deriving explicit MPC formulation of \eqref{eq:optimization1} transforms the given optimization problem into a \textit{multi-parametric Linear Program} (mp-LP) by treating $l_k = \left(x(k),\chi(k),p_c(k)\right) \in \mathcal{P}$ as a parameter vector at current time step $k$, within a predefined feasible polyhedral set $\mathcal{P}\subset\R^4$. The solution of mp-LP gives an explicit MPC control law 
\begin{equation}
    \Delta p_c^\star(l_k) = J_i(l_k) + q_i,
\end{equation}
where $J_i(l_k)\in\R$ and $q_i\in\R$ define a piecewise affine function for all parameter vectors $l_k\in \mathcal{P}_i$ belonging to a polyhedral subspace partition $\mathcal{P}_i\subseteq\mathcal{P}$ of the original set \cite{Bemporad2002}. The number of subspace partitions mostly depends on the number and complexity of constraints, whereas the required offline computational time depends on the length of the prediction horizon.

\section{Centralized Control Design} \label{sec:centralized}

The aim of the centralized grid controller is to provide fast frequency response by manipulating the active power setpoints of all converter-interfaced generators simultaneously. In contrast to decentralized control, which relies only on locally available measurements, an estimate of the dynamical system state can be globally obtained using a wide-area system of Phasor Measurement Units (PMUs), thus improving the regulation accuracy. A benefit of these additional measurements is that FFC can be provided while taking line power flow limits into consideration. In this study we assume the communication links to be reliable and high-speed, therefore neglecting any communication failure scenarios and delays and focusing solely on the underlying control problem.

Similarly to the decentralized MPC, the centralized grid controller is triggered by a large power imbalance. More precisely, PMU measurements at every bus are compared to the scheduled power injections in order to detect the disturbance. Once detected, the power imbalance is used as an input for the MPC problem. Upon activation, the MPC solver runs on a constant clock until average RoCoF values at every bus over a predefined time period fall below a given threshold. The remainder of this section presents the derivation of an appropriate prediction model and formulation of the centralized MPC problem.

\subsection{Simplified System Model}
Following the work in \cite{GrossIREP}, we derive a prediction model that captures frequency dynamics of individual units as well as network line flows, while being simple enough for practical MPC implementation. Each VSC-interfaced unit $i\in\mathcal{N}_c$, where $n_c=\lvert\mathcal{N}_c\rvert$, can be modeled with two dynamic states $x_{c_i} = (\theta_{c_i},\tilde{p}_{c_i})\in\R^2$, reflecting the voltage angle $\theta_{c_i}\in[-\pi,\pi)$ and filtered active power $\tilde{p}_{c_i}\in\R$ from \eqref{eq:w_c}. Using droop control, the angle dynamics can be expressed by
\begin{equation} \label{eq:theta_c}
\dot{\theta}_{c_i} = R_{c_i}^p (\Delta p_{c_i}^{\star}-\tilde{p}_{c_i}),
\end{equation}
thus capturing the frequency response of the converter linearized around a steady-state operating point. 

For synchronous generation, a third-order SG model of the form 
\begin{align}
    M_{s_j}\dot{\omega}_{s_j} &= -D_{s_j}\omega_{s_j} + p_{m_j}^\star - p_{s_j},  \label{eq:swing_eq}\\
    T_{g_j}\,\dot{\tilde{p}}_{s_j} &= -\tilde{p}_{s_j} - K_{g_j} \omega_{s_j}, \label{eq:governor}\\
    \dot{\theta}_{s_j}&=\omega_{s_j} \label{eq:angle}
\end{align}
is employed, where $x_{s_j} = (\theta_{s_j},\omega_{s_j},\tilde{p}_{s_j})\in\R^3$ is the state vector describing the rotor angle $\theta_{s_j}\in[-\pi,\pi)$, rotor speed $\omega_{s_j}\in\R_{\geq0}$, and dynamics of governor control $\tilde{p}_{s_j}\in\R$ of each synchronous generator $j\in\mathcal{N}_g$, with $n_g=\lvert\mathcal{N}_g\rvert$; $p_{s_j}\in\R$ indicates changes in the electrical power output, $M_{s_j}\in\R_{>0}$ and $D_{s_j}\in\R_{>0}$ denote generator inertia and damping constants, whereas $T_{g_j}\in\R_{>0}$  and $K_{g_j}\in\R_{>0}$ represent the governor time constant and control gain respectively. The swing equation \eqref{eq:swing_eq} is linearized around steady state and assumes constant mechanical input $p_{m_j}^\star\in\R_{\geq0}$ over the timescales of interest. A first-order low-pass filter given by \eqref{eq:governor} models the governor dynamics and droop control of the generator \cite{Kundur1994}.

A DC power flow approximation is used to model the network comprising $n_n=\lvert\mathcal{N}_n\rvert$ nodes and $n_b=\lvert\mathcal{N}_b\rvert$ branches, described by the graph Laplacian $L\in\R^{n_n\times n_n}$ (i.e., the bus susceptance matrix of the grid). Under small-signal DC power flow assumptions, the vector $p\in\R^{n_n}$ representing the active power injection at each node can be linearized as
\begin{equation} \label{eq:dc_pf}
    p = L \theta + p_l,
\end{equation}
with $\theta\in\R^{n_n}$ being the vector of nodal voltage angles and $p_l\in\R^{n_n}$ denoting the vector of load power changes at every bus. Line flows $p_b\in\R^{n_b}$ across branches are subsequently computed as $p_b=\hat{X}_bG\theta$, where $\hat{X}_b=\mathrm{diag}(\hat{x}_1^{-1},\dots,\hat{x}_{n_b}^{-1})\in\R^{n_b\times n_b}$ denotes the line susceptance matrix\footnote{$\hat{x}_k$ represents the series reactance of branch $k\in\mathcal{N}_b\subseteq\Z_{\geq0}$.} and $G\in\Z^{n_b\times n_n}$ is the graph incidence matrix.

Finally, a uniform representation of the network comprising $n_n$ nodes, $n_b$ branches, $n_g$ synchronous and $n_c$ converter-interfaced generators can be established by combining \eqref{eq:w_c} with \eqref{eq:theta_c}-\eqref{eq:dc_pf}, resulting in the linear system
\begin{subequations}
\label{eq:ss_centralized}
\begin{align} 
    \dot{x} = \hat{A} x + \hat{B} u,\\
    y = \hat{C} x + \hat{D} u,
\end{align}
\end{subequations}
where the state space matrices $\hat{A}\in\R^{(2n_c+3n_g)\times(2n_c+3n_g)}$, $\hat{B}\in\R^{(2n_c+3n_g)\times(n_c+n_n)}$, $\hat{C}\in\R^{(n_c+n_g+n_b)\times(2n_c+3n_g)}$ and $\hat{D}\in\R^{(n_c+n_g+n_b)\times(n_c+n_n)}$ describe the system, and vectors of variables are defined as
{\small
\begin{subequations}
\begin{align}
    x&=\left(x_{c_1},\dots,x_{c_{n_c}},x_{s_1},\dots,x_{s_{n_g}} \right)\in\R^{2n_c+3n_g}, \label{eq:ss_cen_x}\\
        u&=\left(\Delta p_{c_1}^\star,\dots,\Delta p_{c_{n_c}}^\star,p_{l}\right)\in\R^{n_c+n_n}, \label{eq:ss_cen_u}\\
    y&=\left(f_{c_1},\dots,f_{c_{n_c}},\,f_{s_1},\dots,f_{s_{n_g}}, \,p_{b_1},\dots,p_{b_{n_b}}\right)\in \R^{n_c+n_g+n_b}.\label{eq:ss_net}
\end{align}
\label{eq:vectors}
\end{subequations}
}%
In \eqref{eq:ss_net}, $f_{s_i}=f_b\omega_{s_i}$ and $f_{c_i}=f_b\omega_{c_i}$ represent individual frequencies of SG and VSC units converted into SI, with $f_b=\SI{50}{\hertz}$ being the base frequency. 

\subsection{Centralized MPC Formulation}
The proposed optimization problem resembles the one presented in Section~\ref{sec:decentralized}. Nonetheless, there are few key distinctions, as the centralized controller determines the power output of each converter participating in FFC. The objective function therefore aims at minimizing the total control effort over the full horizon $k\in\mathcal{H}$ and over all converter units $i\in\mathcal{N}_c$:
\begin{subequations}
\label{eq:optimization2}
\begin{alignat}{3}
    & \underset{u}{\min} \quad && \sum_{k\in\mathcal{H}} \sum_{i\in\mathcal{N}_c}  C_{P_i}(k)  \lVert\Delta p_{c_i}^\star(k) \rVert + C_H \left( \lVert\eta_{f} \rVert_\infty +  \lVert\eta_{r} \rVert_\infty \right)\label{eq:objective2}\\
    & \;\mathrm{s.t.} \quad && \forall k\in\mathcal{H}, \forall i \in \mathcal{N}_c, \forall j\in\mathcal{N}, \nonumber \\
    &  && x(k+1) = \hat{A}_d x(k) + \hat{B}_d u(k) \label{eq:states_pred_cen}, \\
    &  && y(k) = \hat{C}_d x(k) + \hat{D}_d u(k) + \begin{bmatrix}
        f_0 \\
        p_{b_0}
    \end{bmatrix}, \label{eq:otputs_pred_cen} \\
    &  && \dot{f}_j(k) = \frac{f_j(k)-f_j(k-1)}{T_s}, \label{eq:rocof_discrete_cen}\\
    &  && p_{c_i}(k) = p_{c_i}^\star + \sum_{r=1}^k\Delta p_{c_i}^\star(r) + R_{c_i}^p (\omega_{c_i}^\star - \omega_{c_i}(k)),  \label{eq:power_estimated_cen}\\
    &  && \chi_i(k+1) = \chi_i(k) + T_s\frac{p_{c_i}^\star-p_{c_i}(k)}{E_{b_i}}, \label{eq:soc_estimated_cen}\\
    &  && \ushort{p}_{c_i,\mathrm{lim}} \leq p_{c_i}(k) \leq \widebar{p}_{c_i,\mathrm{lim}}, \label{eq:powerLimit_cen}\\
    &  && \ushort{\chi}_{i,\mathrm{lim}} \leq \chi_i(k) \leq \widebar{\chi}_{i,\mathrm{lim}}, \label{eq:SoCLimit_cen}\\
    &  && \ushort{f}_{\mathrm{lim}} \leq f_i(k) + R_{c_i}^p \Delta p_{c_i}^\star(k) \leq \widebar{f}_{\mathrm{lim}}, \label{eq:fciLimit_cen}\\
    &  && \ushort{p}_{b,\mathrm{lim}} \leq p_b \leq \widebar{p}_{b,\mathrm{lim}}, \label{eq:branchFlowLimit_cen} \\
    &  && \ushort{f}_{\mathrm{lim}} - \eta_f(k) \leq f_j(k) \leq \widebar{f}_{\mathrm{lim}} + \eta_f(k), \label{eq:frequencyLimit_cen} \\
    &  && \ushort{\Dot{f}}_{\mathrm{lim}} - \eta_r(k) \leq \Dot{f}_j(k) \leq \widebar{\Dot{f}}_{\mathrm{lim}} + \eta_r(k),\label{eq:RoCoFLimit_cen}\\
    &  && \eta_f(k) \geq 0, \eta_r(k) \geq 0, \label{eq:slackConstraints_cen}
\end{alignat}
\end{subequations}
where $\mathcal{N}=\mathcal{N}_g\cup\mathcal{N}_c$ denotes the index set of all generators (including both synchronous and converter-interfaced ones) in the system, and $u(k)\in\R^{n_c+n_n}$ is the vector comprising setpoint changes $\Delta p_{c_i}^\star(k)$ of all VSCs and nodal load injections $p_l(k)$ at time step $k$. The prediction model in \eqref{eq:states_pred_cen}-\eqref{eq:otputs_pred_cen} represents the discrete-time counterpart (denoted by subscript $d$) of the state space given by \eqref{eq:ss_centralized}-\eqref{eq:vectors}, with the vector of load injections $p_l$ in \eqref{eq:ss_cen_u} being populated by PMU measurements of system disturbances and remaining constant throughout the prediction horizon. Vectors $f_0\in\R^{n_c}$ and $p_{b_0}\in\R^{n_b}$ define the linearization point for individual converter frequencies and network line flows. The RoCoF is calculated for all generators in \eqref{eq:rocof_discrete_cen} and branch flows are kept within permissible limits in \eqref{eq:branchFlowLimit_cen}. Constraints on each individual VSC \eqref{eq:power_estimated_cen}-\eqref{eq:fciLimit_cen} are imposed to keep the SoC, power output and frequency spikes within limits, with the notation adapted from \eqref{eq:optimization1}. Frequency and RoCoF constraints are enforced on all generators in \eqref{eq:frequencyLimit_cen}-\eqref{eq:RoCoFLimit_cen}, whereas non-negativity constraints are imposed on slack variables in \eqref{eq:slackConstraints_cen}.

\section{Model Validation and Control Performance} \label{sec:res}

\begin{figure}[t]
    \centering
    \includegraphics[scale=0.8]{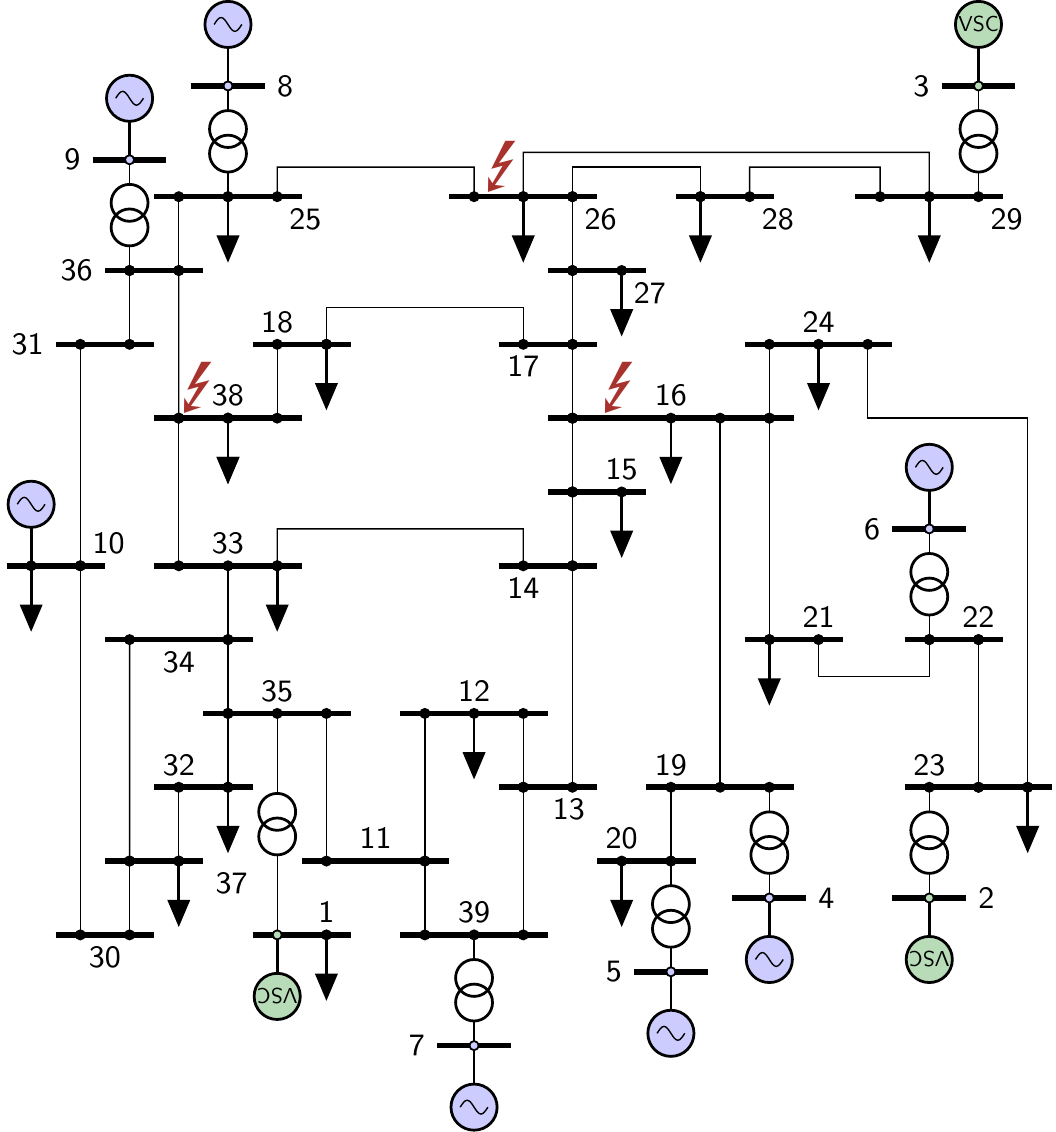}
    \caption{IEEE 39-bus New England test system. Inverter-based generation is placed at nodes $1$, $2$ and $3$. Disturbance locations under consideration are indicated by red symbols.}
    \label{fig:39busIEEE_diag}
\end{figure}

The two proposed FFC schemes have been implemented and evaluated on the IEEE 39-bus test system depicted in Fig.~\ref{fig:39busIEEE_diag}. This is a well-known 10-machine representation of the New England power system, with generator at node 10 representing the aggregation of a large number of generators. The relevant network, load and generation parameters can be found in \cite{Athay1979,39busBook}. The simulations have been performed in \textsc{Matlab} using a DAE model described in \cite{UrosDAE} that encompasses detailed representation of generator and transmission line dynamics. The investigated system comprises seven conventional generators, as three SGs from the original system (precisely at nodes 1, 2 and 3) have been replaced by converter-interfaced units of \SI{1000}{\mega\watt} installed power for the purposes of this analysis. The respective power ratings and output limits of the remaining SGs have been preserved. All VSCs operate in grid-forming mode and are equipped with the supervisory FFC layer.

The disturbances are generated through step changes in active power at network buses of interest, thus emulating either a loss of generator or a loss of load. In this study we assume that the first stage of automatic load-shedding is initiated in case of frequency deviation beyond $\pm\SI{0.5}{\hertz}$, whereas the RoCoF protection is triggered at $\pm1\,\mathrm{Hz/s}$ for RoCoF measurements averaged over a \SI{250}{\milli\second} cycle. Therefore, the frequency-related thresholds in \eqref{eq:optimization1} and \eqref{eq:optimization2} are set as follows: $\ushort{f}_{\mathrm{lim}}=\SI{49.5}{\hertz}$, $\widebar{f}_{\mathrm{lim}}=\SI{50.5}{\hertz}$, $\ushort{\Dot{f}}_{\mathrm{lim}}=-1\,\mathrm{Hz/s}$ and $\widebar{\Dot{f}}_{\mathrm{lim}}=1\,\mathrm{Hz/s}$. The battery SoC and VSC power output are defined in per unit, and hence the minimum and maximum limits are set to $0$ and $1$, respectively.

The prediction horizon of the MPC-based controller is set to three time steps with a sampling period of \SI{250}{\milli\second}. On the one hand, the prediction horizon length of \SI{750}{\milli\second} reflects a trade-off between controller performance and computational effort. On the other hand, the MPC sampling period is selected such that it exceeds all delays associated with the converter and supervisory layer, as well as the time needed to compute the optimal control decisions. The instantaneous RoCoF estimate $\dot{\omega}_\mathrm{max}$ required for computation of the disturbance magnitude $\Delta P$ is obtained by averaging the the internal RoCoF state signal $\dot{\omega}$ over a time interval of \SI{10}{\milli\second} in the immediate aftermath of a disturbance.

The following sections will first analyze the impact of parametrization on the accuracy of the frequency prediction model in \eqref{eq:ss_general}, and compare it against the RoCoF-based approach in \eqref{eq:papangelis_df}. Subsequently, the controller performance for different disturbance locations and magnitudes is evaluated, followed by a discussion on the battery storage requirements and explicit MPC formulation. 

\subsection{Prediction Model Validation}

\begin{figure}[!t]
    \centering
    \scalebox{1.125}{\includegraphics{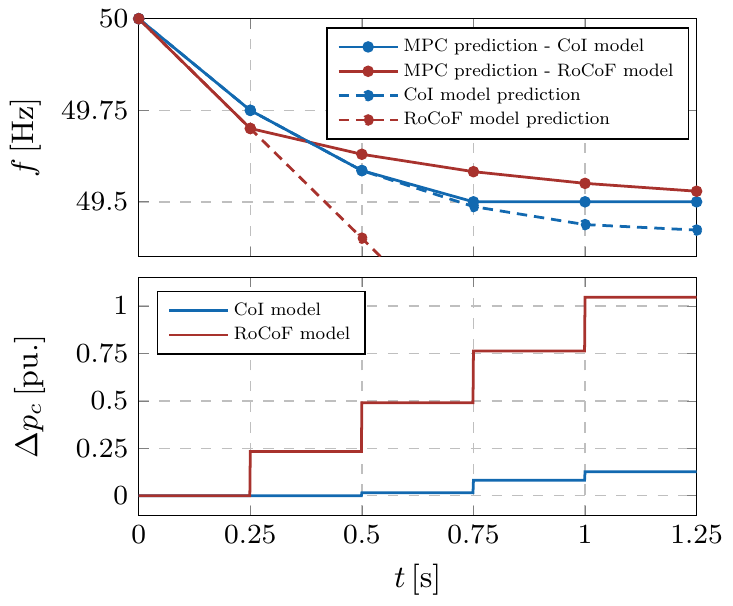}}
    \caption{Comparison of decentralized MPC performance under two frequency prediction models. Anticipated frequency evolution and adjusted power injections are used for evaluation.}
    \label{fig:predComp}
\end{figure}

The controller operation for $N=5$ time steps is shown in Fig.~\ref{fig:predComp}, comparing the performance of the proposed frequency prediction model against the RoCoF-based one. Predicted frequency evolution in case of no corrective actions (indicated by the dashed lines) demonstrates the conservative nature of the RoCoF-based approach. In particular, due to constant RoCoF estimate throughout the whole prediction horizon, the anticipated frequency nadir is well below the actual value. As a result, the control effort is significantly higher than with the CoI model. Being proven advantageous and more efficient, only the CoI model is considered hereinafter.

\begin{figure}[!b]
    \centering
    \scalebox{1.125}{\includegraphics{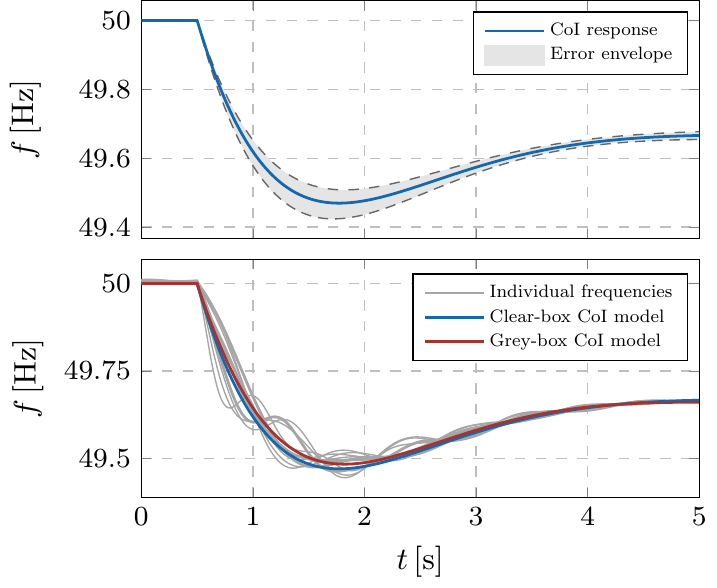}}
    \caption{Accuracy of the frequency prediction model: sensitivity to model parameters (top) and CoI-model verification (bottom).}
    \label{fig:envAndGrey}
\end{figure}

It was noted previously that inaccurate knowledge of system parameters in the CoI model could lead to degradation of response quality and potential control failure. To investigate the severity of this problem, a parametric sensitivity analysis was performed for an arbitrary disturbance by considering all possible combinations of two SGs in the IEEE 39-bus system going offline. The error envelope around the median frequency response, derived from simulations and illustrated in the upper plot of Fig.~\ref{fig:envAndGrey}, indicates the maximum nadir error of $\approx\SI{0.01}{\percent}$ for the considered generator sets.

The aforementioned issue with parameter uncertainty can be mitigated through the grey-box system identification procedure outlined in Section~\ref{sec:decentralized}. To illustrate the efficiency of such approach, an active power step change of \SI{1575}{\mega\watt} at bus $16$ is simulated, with generator frequencies presented in Fig.~\ref{fig:envAndGrey}. The individual frequencies are compared to the frequency prediction of the CoI model, once parametrized using the exact generator parameters (clear-box) and once through the grey-box model identification procedure. The individual generator frequencies are matched well by the CoI model response in both cases, with a negligible difference between the two parametrization methods. The data used for system identification process were retrieved by simulating a different load step disturbance and collecting VSC frequency measurements at node $1$. Parameter fitting was subsequently employed using the \textsc{Matlab} System Identification Toolbox with an RMSE of \SI{2}{\percent}.
    
\subsection{Control Performance and Comparison}

\begin{figure}[!b]
    \centering
    \scalebox{1.125}{\includegraphics{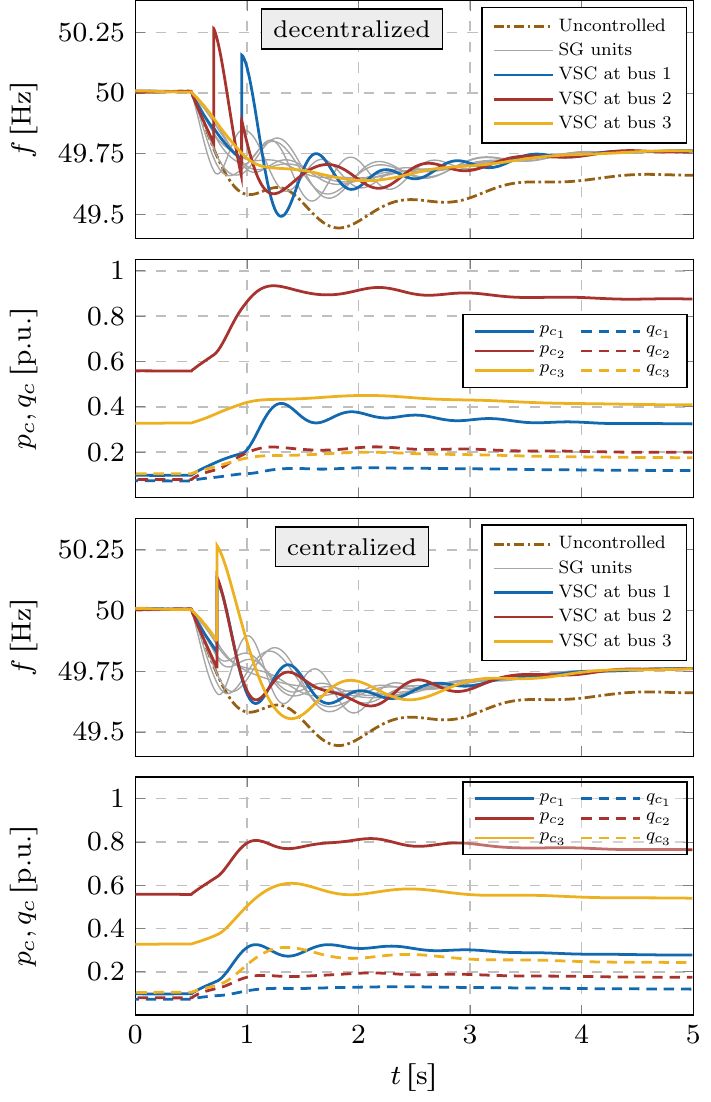}}
    \caption{Individual frequency, active and reactive power output responses for the decentralized (top) and centralized (bottom) FFC scheme following a disturbance at bus $16$. Dashed line depicts the worst-case generator frequency without the use of FFC.}
    \label{fig:bus16_comparison_f}
\end{figure}

Performance of the decentralized control depends on how well the power imbalance can be estimated through internal RoCoF state of the active power controller. It is well known that frequencies, and correspondingly the instantaneous RoCoF values, will vary significantly at different nodes of a large power system following a disturbance. For this reason, a symmetrical placement of converters providing FFC support is crucial for coverage of disturbances at as many system nodes as possible. Since power ratings of all VSCs are the same, each unit is expected to participate equally in disturbance mitigation. Unlike the decentralized approach, the centralized grid controller obtains an accurate disturbance estimate through wide-area measurements, independent of the disturbance location, and hence operates with low error margins.

\begin{figure}[!t]
    \centering
    \scalebox{1.125}{\includegraphics{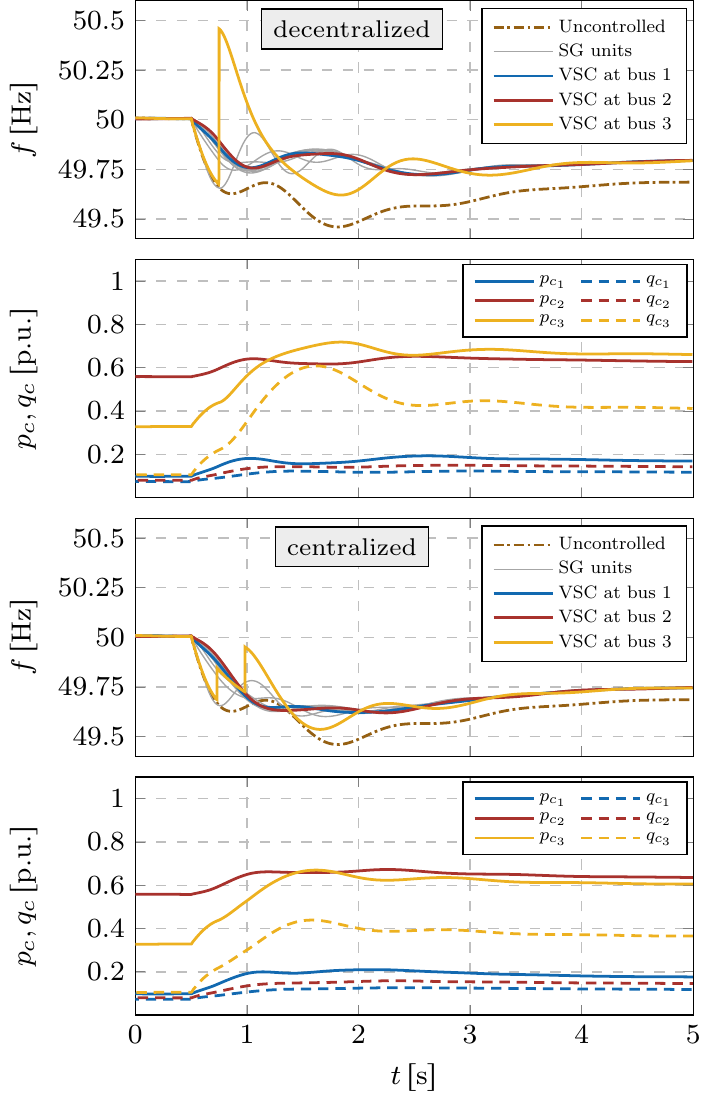}}
    \caption{Individual frequency, active and reactive power output responses for the decentralized (top) and centralized (bottom) FFC scheme following a disturbance at bus $26$.}
    \label{fig:bus26_comparison_f}
\end{figure}

In the following, we evaluate and compare the performance of both control approaches by analyzing the system response for different disturbance locations indicated in Fig.~\ref{fig:39busIEEE_diag}. The values of applied and estimated disturbance magnitudes for every considered bus and for each VSC are presented in Table~\ref{tab:bus_disturbances}.

\begin{table}[!b]
\renewcommand{\arraystretch}{1}
\caption{Fault scenarios at different buses with indicated applied disturbance magnitudes and estimated imbalances for each VSC located at nodes $1$, $2$ and $3$, respectively.}
\label{tab:bus_disturbances}
\noindent
\centering
    \begin{minipage}{\linewidth} 
    \renewcommand\footnoterule{\vspace*{-5pt}} 
    \begin{center}
    \scalebox{1}{%
        \begin{tabular}{ c | c | c | c | c }
            \toprule
            \multirow{2}{*}{\textbf{Bus}} & \multirow{2}{*}{\textbf{Disturbance $\boldsymbol{[\mathrm{MW}]}$}} & \multicolumn{3}{c}{\textbf{Estimated disturbance $\boldsymbol{[\mathrm{MW}]}$}}\\ 
            \cline{3-5}
            & & VSC 1 & VSC 2 & VSC 3 \\
            \cline{1-5}
            $16$ & $1575$ & $1550$ & $1955$ & $835$\\
            $26$ & $1430$ & $1100$ & $790$ & $3860$\\
            $38$ & $1850$ & $1390$ & $520$ & $650$\\
            \arrayrulecolor{black}\bottomrule
        \end{tabular}
    }
        \end{center}
    \end{minipage}
\end{table}

First, let us consider a power disturbance of \SI{1575}{\mega\watt} at node $16$. Fig.~\ref{fig:bus16_comparison_f} shows frequencies, active and reactive power outputs of individual generators for both FFC approaches. The dashed line represents the lowest (i.e., ``worst-case'') frequency nadir of any unit in the system when FFC is disabled. Note that the droop control of all units is still active. The VSC at node $3$ remains inactive due to the large electrical distance to the fault location and consequent underestimation of the disturbance. However, the support from the other two converters is sufficient to compensate the disturbance and prevent load-shedding. On the other hand, the global MPC-based grid controller dispatches all three units equally, with the identical total control effort for both MPC approaches.

\begin{figure}[!t]
    \centering
    \scalebox{1.125}{\includegraphics{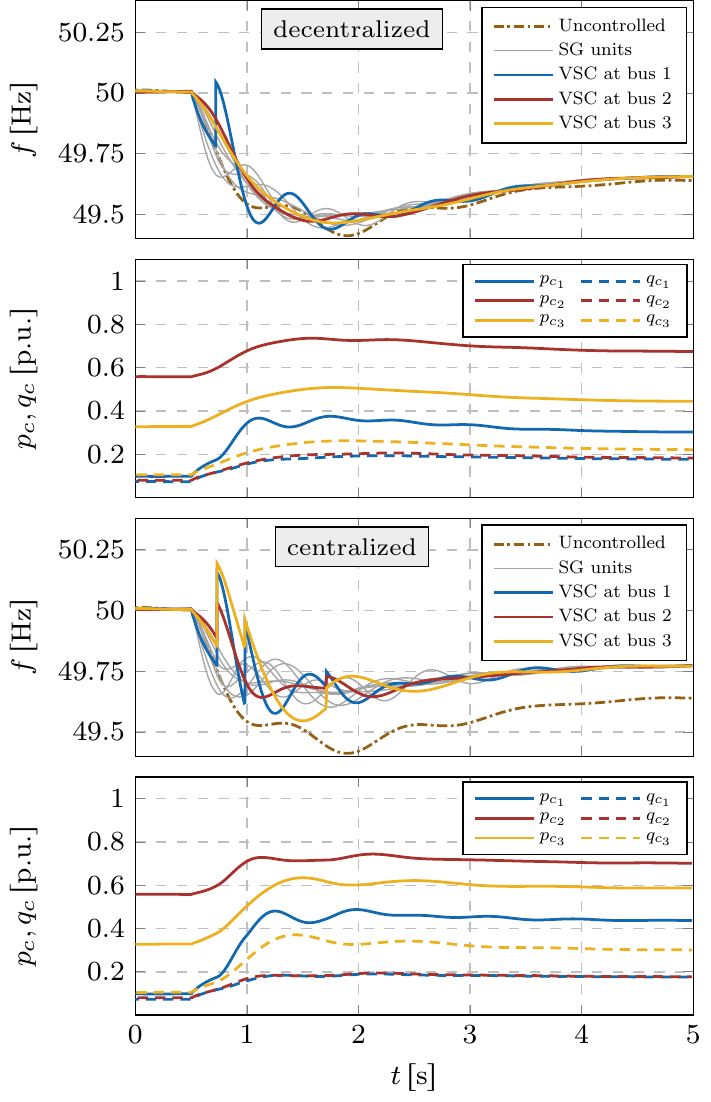}}
    \caption{Individual frequency, active and reactive power output responses for the decentralized (top) and centralized (bottom) FFC scheme following a disturbance at bus $38$.}
    \label{fig:bus38_comparison_f}
\end{figure}

In contrast, Fig.~\ref{fig:bus26_comparison_f} illustrates the control performance for a disturbance at bus $26$, in the vicinity of converter-interfaced DG at node $3$ . Hence, this VSC unit overestimates the disturbance and significantly increases its power output, whereas the other two converters remained idle. Similarly, the centralized controller increases only the power output of the VSC at node $3$, which due to its location has the most influence on the relevant frequency dynamics. An overall lower control effort is employed in the centralized approach.

Lastly, we consider a disturbance at bus $38$, located such that it exhibits a large electrical distance between all three VSCs. The individual frequency response of all generators is given in Fig.~\ref{fig:bus38_comparison_f}. Understandably, decentralized controllers underestimate the disturbance due to a large electrical distance from the fault location. VSC-based DG at bus $1$ is the only one to react, though insufficiently to compensate for the disturbance and prevent load-shedding. Having accurate global measurements, the centralized grid controller detects the disturbance and reacts appropriately and timely through all available converter-interfaced units.

\subsection{Analysis of DC-side Dynamics}
\begin{figure}[!b]
    \centering
    \scalebox{1.125}{\includegraphics{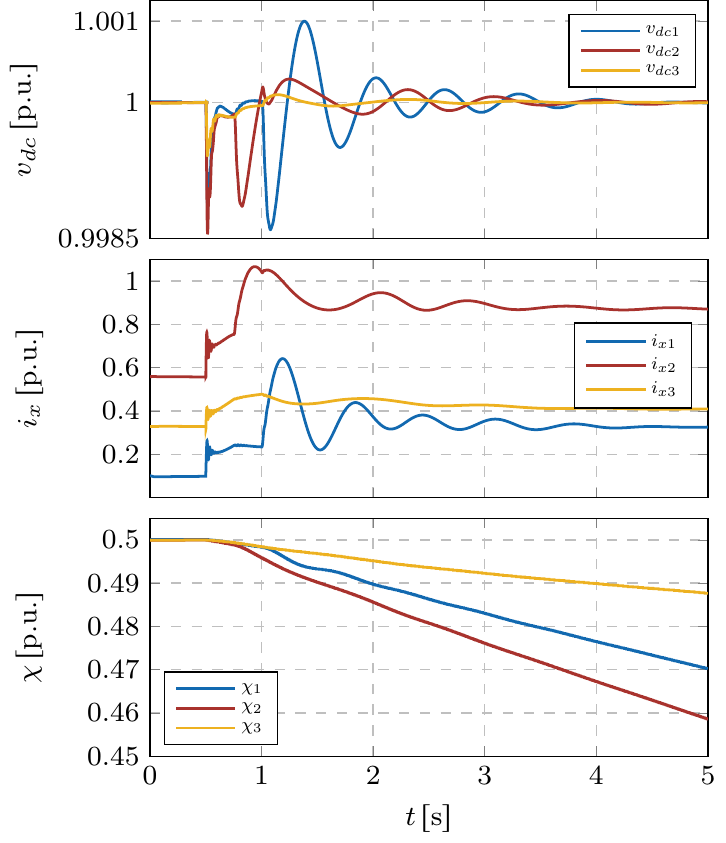}}
    \caption{Individual DC voltage, input current and SoC responses for the decentralized FFC scheme following a disturbance at bus $16$.}
    \label{fig:bus16_dcside_dynamics}
\end{figure}

The most relevant variables describing the dynamics of the DC-side circuit, namely the capacitor voltage $v_\mathrm{dc}$, the input current $i_x$ and the battery SoC $\chi$ are shown in Fig.~\ref{fig:bus16_dcside_dynamics} for individual VSC units and the disturbance at bus $16$. The installed energy capacity of the batteries is assumed to be \SI{10}{\MWh}, with the initial SoC for all three inverters set to $0.5 \,\mathrm{p.u.}$ The disturbance and the inverter setpoint changes cause power imbalances at the capacitor node, leading to DC voltage dips which are quickly restored by the DC-side controls using available energy of the battery. As a consequence, SoC levels of individual inverters decrease at different rates depending on the applied setpoint change by the FFC layer. Note that for a smaller battery size the SoC levels would decrease faster, justifying inclusion of the SoC constraints in the MPC problem formulation. Finally, the input DC current indicates faster dynamics compared to the output power (see Fig.~\ref{fig:bus16_comparison_f}), which could potentially lead to high current injections. While highly relevant for safe operation of power electronic devices, such problems and limitations are addressed by the overcurrent protection schemes incorporated within the device-level control, and are therefore out of the scope of this work.

\subsection{Analysis of Computational Efficiency}
The solution of the linear MPC optimization problems \eqref{eq:optimization1} and \eqref{eq:optimization2} was performed using the CPLEX solver, an LP solver based on interior point algorithms, for numerical computation and YALMIP \cite{yalmip} for high-level modeling. The computational time required for solving the decentralized MPC problem is \SI{185}{\milli\second}, whereas the centralized MPC problem is solved in \SI{343}{\milli\second} on average. The computational efficiency for the decentralized controller can further be improved by generating an explicit solution, as discussed in Sec.~\ref{sec:decentralized}. The explicit MPC solution was generated using YALMIP for modeling and the MPT3 toolbox \cite{MPT3} for low-level numerical solution of the multi-parametric optimization problem. The solution partitions the parameter space in $452$ regions and takes \SI{95}{\second} to be generated. In this case, the time required to obtain the optimal control inputs reduces to \SI{15.86}{\milli\second}.

\section{Conclusion} \label{sec:conclusion}
This paper presents a novel FFC scheme for converter-interfaced DGs in low-inertia systems, which exploits their fast response to prevent load-shedding scenarios. An MPC-based supervisory control layer is added to the traditional converter control scheme, which in response to a large disturbance manipulates converter setpoints to contain the frequency within predefined bounds. Both centralized and decentralized control approaches were considered and compared. Novel prediction models were developed and compared to the state-of-the-art, which showcased improvements in prediction accuracy. 

The centralized approach has proven to be advantageous in several cases indicating the value of fast communication infrastructure. On the other hand, the decentralized controller proves to be efficient for containing frequency excursions for disturbances occurring in the vicinity of at least one or few converter-interfaced generators. Advantages of this approach are a simple, plug-and-play architecture, as well as a low-cost and computationally efficient implementation. Simulation results do not suggest any potential frequency instabilities arising from the control actions of the MPC-based supervisory layer, but we leave a theoretical proof of stability for future work.

\subsection{Outlook and Future Work} \label{subsec:outlook}

An interesting avenue for future work is a distributed approach, where controllers of individual VSCs carry out their calculations in separate processors, but efficiently cooperate using only communication links between different local controllers. The extension of this study will focus on applying machine learning frameworks such as data-driven support vector machine on the CoI model to improve the system frequency prediction in the aftermath of a disturbance.

Present work focuses on designing the FFC layer that aims to react and stabilize the system in the events of generator outages and sudden load changes. As the RoCoF estimates are an essential part of the proposed controller, it is important to consider the impact of transmission line faults on the control behavior. Such events as well as the instances of converter disconnection will be the subject of future research.

Upon the successful frequency containment by virtue of joint efforts of the primary and fast frequency control schemes, the AGC is activated and slow-acting reserves are dispatched to replace the missing generation. The converter setpoints can now be readjusted (i.e., decreased) and a recovery period whose aim is to prepare the FFC providing units for the next operation cycle begins. Development of optimal schemes for FFC deactivation and recovery will be addressed in future work.  



\bibliographystyle{IEEEtran}
\bibliography{bibliography}

\end{document}